% ...

\documentstyle[psfig,newpasp,twoside]{article}

\newcommand{\msun}{\mbox{${\rm M}_\odot$}}
\newcommand{\rsun}{\mbox{${\rm R}_\odot$}}
\newcommand{\nbody}{\mbox{{{\em N}-body}}}
\newcommand{\tcrss}{\mbox{${t_{\rm hm}}$}}
\newcommand{\trlx}{\mbox{${t_{\rm rlx}}$}}

\newcommand{\rcore}{\mbox{${r_{\rm core}}$}}

\newcommand{\rvir}{\mbox{${r_{\rm vir}}$}}

\newcommand{\rhm}{\mbox{${r_{\rm hm}}$}}
\newcommand{\rhocore}{\mbox{${\rho_{\rm core}}$}}
\newcommand{\kms}{\mbox{${\rm km~s}^{-1}$}}
\def\unit#1{{\mbox{[{\rm #1}]}}}
\def\apgt{\ {\raise-.5ex\hbox{$\buildrel>\over\sim$}}\ }
\def\aplt{\ {\raise-.5ex\hbox{$\buildrel<\over\sim$}}\ }

\begin{document}
\title{Collisions in compact star clusters}

\author{Simon F.\ Portegies Zwart}
\affil{Department of Astronomy,
		  Boston University,
		  725 Commonwealth Ave.,
		  Boston, MA 02215, USA \\
		Hubble Fellow}

\date{received; accepted:}
\markboth{Portegies Zwart: collisions in clusters}{}

\begin{abstract}
The high stellar densities in young compact star clusters, such as the
star cluster R\,136 in the 30 Doradus region, may lead to a large
number of stellar collisions.  Such collisions were recently found to
be much more frequent than previous estimates.  The number of
collisions scales with the number of stars for clusters with the same
initial relaxation time.  These collisions take place in a few million
years.  The collision products may finally collapse into massive black
holes.  The fraction of the total mass in the star cluster which ends
up in a single massive object scales with the total mass of the
cluster and its relaxation time. This mass fraction is rather
constant, within a factor two or so.  Wild extrapolation from the
relatively small masses of the studied systems to the cores of
galactic nuclei may indicate that the massive black holes in these
systems have formed in a similar way.
\end{abstract}


\section{Introduction}
In the central regions of dense star clusters and galaxy bulges
collisions between stars are rather common.  Such collisions are
likely to play an important role in the formation of exotic objects
such as blue stragglers (Sanders 1970; McNamara \& Sanders
1976)\nocite{san70}\nocite{1976A&A....52...53M}, X-ray binaries
(Fabian et al.~1975)\nocite{fpr75} and millisecond pulsars (Lyne et
al.\ 1987;
1988),\nocite{1987Natur.328..399L}\nocite{1988Natur.332...45L} but
have also an important influence on the dynamics of these clusters
(Quinlan \& Shapiro 1990; Quinlan et al.~1995, Lee
1995).\nocite{1995ApJ...440..554Q}\nocite{1995MNRAS.272..605L}

Both the stars and the parent cluster evolve on comparable time
scales, and cluster dynamics and stellar evolution are quite closely
coupled.  For example, massive stars tend to segregate to the core due
to dynamical friction, increasing their collision probability.
Collision products are even more massive, leading to the possibility
of runaway merging, if the collision rates can remain high enough in
the few Myr before the stars explode as supernovae (Lee 1987; Quinlan
\& Shapiro 1990).\nocite{1987ApJ...319..801L}
\nocite{1990ApJ...356..483Q} However, these rates are determined by
the dynamical state of the cluster, which in turn is strongly
influenced by stellar mass loss and binary heating.  The only way to
treat this intimate coupling between stellar collisions and cluster
dynamics is to perform \nbody\ simulations in which the stars are
allowed to evolve and collide with one another in a fully
self-consistent way.

Recently Portegies Zwart et al.\, (1999, hereafter PZMMH) demonstrated
via direct \nbody\, calculations of dense star clusters that multiple
collisions are rather common and that the collision products may grow
to very high masses within a few million years. In the models the
collision rate is two orders of magnitude higher than semi-analytic
estimates indicate. Although the calculations are still rather
limited, excluding the effects of primordial binaries and for a
limited number of stars, they argue that the results could be scaled
with respect to the total mass and initial relaxation time, i.e., with
$N$. 

In this paper we report the results of a series of \nbody\ simulations
modeling young and compact star clusters, such as R\,136 in the
30\,Doradus region in the Large Magellanic Cloud.  This cluster is
particularly interesting because a strong coupling between stellar
evolution and stellar dynamics may exist. In addition to this,
excellent observational data is available.  Many unusually bright and
massive stars (e.g.\ Massey \& Hunter 1998)\nocite{MH98} are present
in R\,136 which, due to the high central density of $10^6$ to $10^7$
stars pc$^{-3}$, are likely to interact strongly with each other.

We find that runaway collisions of massive stars can occur, and that
the most massive star grows in mass through merging with other stars
until it collapses to a black hole.  The growth rate of this star is
much larger than estimates based on simple cross-section arguments,
because the star is typically found in the cluster core, and tends to
form binaries with other massive stars.  The maximum mass of the
runaway collision product scales with the total mass of the star
cluster and the scaling factor is rather constant.  After half an
initial relaxation time between 0.9\% and 2.4\% of the total mass of
the star cluster ends up in a single object.

\section{Methods}
Our findings are based on series of direct \nbody\, simulations with
the ``Starlab'' software environment (Portegies Zwart et al.\, 1999:
see {\tt http::/www.sns.ias.edu/$^\sim$starlab}), using the
special-purpose computer GRAPE-4 to speed up the calculations (Makino
et al.\,1997).  The evolution of stars is taken into account self
consistently with the dynamical evolution of the star cluster.  A
collision is assumed to occur when two stars ($i$ and $j$) approach
each other within a distance $d = 2 (r_i + r_j)$, where $r_i$ and
$r_j$ are the radii of the stars involved.  A collision between two
main-sequence stars with masses $m_i$ and $m_j$ results in a single
rejuvenated main-sequence star with mass $m_i + m_j$.  Smooth-particle
hydrodynamic simulations of collisions between main-sequence stars
indicate that at maximum a few percent of the total mass is lost (see
e.g.: Lai et al.\ 1993; Lombardi et al.\ 1995; 1996).  Consequently,
mass loss during the merger event is ignored.
\nocite{1993ApJ...412..593L}\nocite{1996ApJ...468..797L}\nocite{1995ApJ...445L.117L}

\section{Selection of initial conditions}\label{Sect:Init}
We selected the initial parameters for the models to mimic a class of
star clusters similar to the Galactic cluster NGC 3606 or the young
globular cluster NGC 2070 (R136) in the 30 Doradus region of the
Large Magellanic Cloud. 

\subsection{The star cluster R\,136 in the 30\,Doradus region}
The half-mass radius ($r_{\rm hm}$) of R\,136 is about 1 parsec
(Brandl et al.\ 1996),\nocite{BSB+96} and the core radius $r_c \simeq 
0.02$\,pc (Hunter et al.\ 1995).\nocite{1995ApJ...444..758H} The total
mass $M \sim 2\,\times\, 10^4\,$\,\msun.  With an assumed mean mass of
0.6\,\msun\, the cluster thus contains about 35\,000\, stars.  The
corresponding central density is of the order of
$10^6$\,\msun\,pc$^{-3}$.  The age of R\,136 is $\sim 3$--4\,Myr
(Campbell et al.\ 1992).\nocite{CHH+92} The Galactic star cluster NGC 3606 is
somewhat smaller in size and its total mass is larger resulting in a
denser core (Moffat et al.\ 1994; Drissen et
al.\,1995).\nocite{1994ApJ...436..183M}\nocite{1995AJ....110.2235D}

For both clusters, the tidal effect of their parent galaxy (for R\,136
that is the Large Magellanic Cloud) is small and we therefore 
neglected the effect of a tidal field.

\subsection{Scaling the dynamical time scale}
The evolution of an isolated star cluster is driven by two-body
relaxation. Therefore, we set up the initial model so that it has the
same relaxation time scale as the real cluster.  The relaxation time is
calculated with
\begin{equation}
	t_{\rm rlx} \propto {N \over \ln (\gamma N)} \tcrss.
\label{Eq:trlx}\end{equation}
Here $N$ is the number of stars and $\gamma$ is a scaling factor,
introduced to model the effects of the cut-off in the long range
Coulomb logarithm (see Giertz \& Heggie 1996;
1994).\nocite{GH96}\nocite{1994MNRAS.268..257G} Here \tcrss\ is the
half-mass crossing time of the cluster is
\begin{equation}
\tcrss \simeq 57 \left( {[\msun] \over M} \right)^{1/2}
                 \left( {\rhm \over [{\rm pc}]}\right)^{3/2} \, \unit{Myr}.
\label{Eq:thc}\end{equation}
Here $r_{\rm hm}$ is its half mass radius.  

The radius of the scaled model $r_{\rm model}$ is then computed by 
substitution of Eq.\,(\ref{Eq:trlx}) into Eq.\,(\ref{Eq:thc}) and
for simplicity we neglect the logarithmic term:
\begin{equation}
	r_{\rm model} \propto \left( { N_{\rm real} \over N_{\rm model} } 
   		              \right)^{1/3}
			r_{\rm real}.
\label{Eq:rmodel}\end{equation}
Here $r_{\rm real}$ is the radius of the real cluster.  We decided to
use $\gamma =1$.  

\subsection{Scaling the collision cross section}\label{Sect:collscale}
We want the model clusters to have the same collision rate per star as
the real cluster.  Scaling the initial conditions to assure that the
model cluster has the same relaxation time causes it to be larger than
the real system (Eq.\,\ref{Eq:rmodel}).  The correct collision rate
per star is then obtained by scaling the sizes of the stars
themselves.

The number of collisions per star per unit time is given by
(Spitzer 1987)\nocite{spi87} 
\begin{equation}
 	n_{\rm coll} \propto n_c \sigma v.
\label{Eq:Ncoll}\end{equation}
Here $n_c$ is the number density of the stars in the core, $\sigma$ is
the collision cross section (for approach within some distance $d$),
and $v$ is the velocity dispersion.  These are given by the following
proportionalities:
\begin{eqnarray}
	n_c 	&\propto& 	{N \over r_c^3}, \nonumber	\\
	\sigma 	&\propto& 	d^2 + {d \over v^2}. 
\label{Eq:crossection}\end{eqnarray}
We neglect the $d^2$ term in the cross section.
Expressed in real units and assuming scaling according to
Eq.\,(\ref{Eq:rmodel}), we may write
\begin{eqnarray}
	{N \over r_c^3} &\propto& N^2, \nonumber	\\
	v 	        &\propto& 	N^{2/3}.
\end{eqnarray}
The number of collisions then becomes
\begin{equation}
 	n_{\rm coll} = d N^{4/3}. 
\label{Eq:ncoll}\end{equation}
The distance at which a collision occurs therefore scales as
\begin{equation}
 	d \propto N^{-4/3}. 
\end{equation}

\subsection{The models}
We performed runs with 24k, 12k, 6k and 3k stars ($1{\rm k} \equiv
1024$).  All runs start from the same initial relaxation time and
density profile.

All simulations start at $t=0$ by assigning masses of stars between
0.1\,\msun\ and 100\,\msun\ from the mass function suggested for the
Solar neighborhood by Scalo (1986).\nocite{sca86}

The initial density profile and velocity dispersion are taken from a
King (1966) model with $W_0 = 6$.\nocite{kin66} We chose $\rvir =
0.31$\,pc for the models with 12k stars. This results in a core
radius $\rcore \approx 0.072$\,pc and a core density of $\rhocore \sim
3.6\,\times 10^{5}\,\msun\,{\rm pc}^{-3}$. The central velocity
dispersion for these models is about 8.7\,\kms\ and the initial
half-mass relaxation time $\trlx \sim 10$\,Myr.
All computations were continued until $t=10$\,Myr.

\section{Results}

\subsection{Mass segregation}

The models are characterized by an initial expansion; the core radius
increases from 0.07 to 0.18\,pc within the first 4 million years
(roughly 50 crossing times). Together with the expansion of the
cluster, the density in the core decreases by about an order of
magnitude.

The expansion of the cluster is driven mainly by the formation and
subsequent heating of binaries.  The cluster is still too young for
significant stellar evolution to occur, therefore mass loss by
stellar evolution is relatively unimportant. The first binaries are
formed shortly after the start of the simulation. One of these
binaries typically contains two of the most massive stars.

Mass segregation has brought these two stars into the core on a time
scale similar to the crossing time of the cluster. On their first
orbit through the cluster core these stars simply did not leave but
stayed in the core. This is not entirely unexpected as the most
massive objects are between 50 and 100 times more massive than the
mean mass in the stellar system. For the selected initial conditions
the time scale for mass segregation is therefore expected to be of the
order of a crossing time.

Further mass segregation results in an increase in the mean mass in
the cluster core from the initial 0.6\,\msun\, to double this value
within a million years. After that the mean mass in the core stays
more or less constant.

\subsection{The collision rate}\label{Sect:runaway}

In all our model calculations we observe a collision rate which is
about two orders of magnitude higher than expected on cross section
arguments.

The rate at which stars in a cluster experience collisions can be
estimated via Eq.\,\ref{Eq:Ncoll}.  Following the derivation by
Portegies Zwart et al.\, (1997) in which a Maxwellian velocity
distribution with velocity dispersion $v$ and the gravitational
focusing cross section from Eq.\,\ref{Eq:crossection} are adopted, the
number of collisions in the cluster per Myr is expressed as (see their
Eq.\,14)
\begin{equation}
	\Gamma \approx  \left(\frac{n_c}
			       {10^4 {\rm pc}^{-3}}\right)^2 
		            \left(\frac{r_c}{{\rm pc}}\right)^3
			    \left(\frac{ m }{\msun}\right) 
		           \left(\frac{d}{\rsun}\right) 
		            \left(\frac{\kms}{ v }\right)
			\;\; [{\rm Myr}^{-1}].
\label{Eq:rate}
\end{equation}
Here $m$ is the mean stellar mass.

For the $N=12$\,k runs this results in about 0.3 collision/Myr, or
about 3 collisions during the entire simulation, assuming that the
cluster parameters do not change in time. As we discussed in the
previous section, the core density in fact drops by about an order of
magnitude during the first 4\, million years.  If we take this effect
into account, the expected number of collisions is less than unity.
The actual number of collisions in each simulation exceeds 10 (for the
models with 12k stars).  The major cause of this large discrepancy is
the formation of binaries and mass segregation.

\begin{figure}[ht]
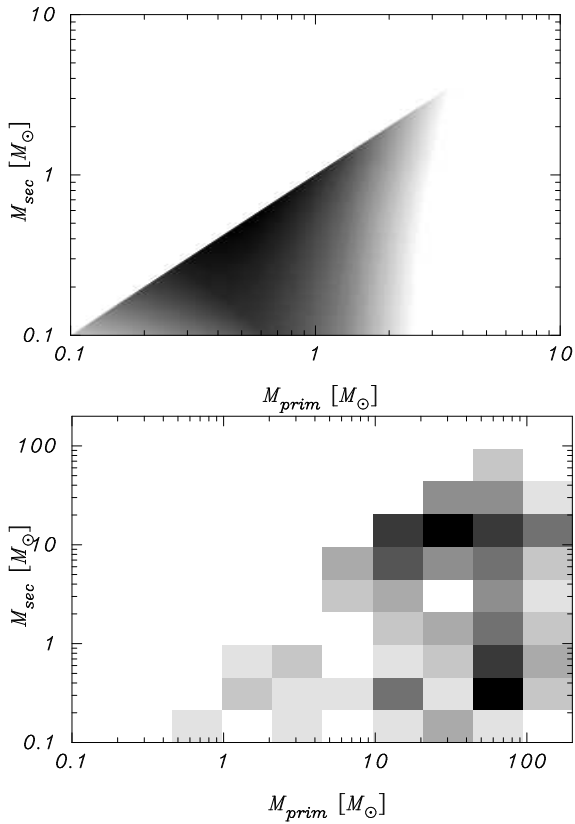

\psfig{figure=simon_1a.ps,width=7.5cm,angle=-90}
\psfig{figure=simon_1b.ps,width=7.5cm,angle=-90}
\caption[]{Relative collision rates between a primary and a secondary
star as expected from Eq.\,\ref{Eq:rate} (top panel) and resulting
from the calculations (lower panel).  The shading in linear in the
encounter probability for the top panel and in the number of
collisions for the lower panel.  Darker shades indicate higher rate.
Since the probability distribution is symmetric line the axis of equal
mass, only the lower half is displayed.}
\label{fig:Renc}
\end{figure}

The importance of mass segregation is illustrated in
Fig.\,\ref{fig:Renc}, which gives the theoretical probability
distribution for stars with mass $m_{\rm prim}$ to collide with
lower-mass stars of mass $m_{\rm sec}$ in the top panel and the
distribution of collisions actually observed in our simulations in the
bottom panel.

The cross section arguments imply that low mass ($m\aplt 1$\,\msun)
stars are most likely to collide.  In our simulations, however,
high-mass stars predominantly participated in encounters.  The most
massive star typically participates in numerous collisions with other
stars. The mass of this runaway grows to exceed 120\,\msun and the
rejuvenation of the runaway merger delays its collapse to a compact
object following a supernova.  Such a star could be visible in the
core of young star clusters as a blue straggler.

The reason for the discrepancy between the formal cross-section
arguments and the results of our simulations is the neglect of mass
segregation and binary formation in the former estimates.  In the
simulations the most massive stars sink to the core due to dynamical
friction within a few half-mass crossing times, and form close
binaries by 3-body interactions.  The larger cross section of these
binaries increases the collision rate and makes them favored
candidates for encounters.

\subsection{Black holes in dense star clusters}
When the runaway merger collapses to a black hole it is typically a
member of a rather close binary.  Upon dissociation of the binary, the
black hole is ejected from the core, but not from the cluster.  Since
the compact object is still considerably more massive than average,
mass segregation brings it back in the core within a few crossing
times (see e.g.\,Hut, McMillan, \& Romani 1992). \nocite{hmr92} New
close binaries can be formed once the black hole has returned to the
core of the star cluster.  After an episode of hardening the binary
may become visible as an X-ray source when the companion star starts
to transfer mass to the black hole.  Such a high-mass X-ray binary
should be easily observable by X-ray satellites. The age at which such
a binary can form is at least $\sim 4$\,Myr, the minimum time needed
for a black hole to form.  It is likely to take considerably longer
because the black hole has to return to the core after its ejection.

The star cluster R\,136 is therefore ``too young'' for such a binary
to exist.  The star Mk\,34 at a distance of about 2.5\,pc from the
center of R\,136, however, is associated with a persistent X-ray
source with a luminosity of $\sim 10^{36}$\,erg\,s$^{-1}$ (Wang
1995).\nocite{1995ApJ...453..783W} Wang suggests that the binary
contains a black hole of between 2.4 and 15\,\msun\ accreting from the
dense wind of its spectral type WN4.5 Wolf-Rayet companion.  This
star, Mk\,34, can be classified as a ``blue straggler,'' as its
estimated age is about 1\,Myr, considerably smaller than the age of
the cluster (De Marchi et al.\ 1993).\nocite{DMNA+93}

Because R\,136 is too young for such an X-ray binary to be formed from
two collision products, it most likely formed from a primordial binary
ejected from the cluster core following the supernova which formed the
black hole.

\subsection{Black holes in central star clusters}
The centers of galactic nuclei contain supermassive black holes and it
is still unclear how they form (Rees 1999). The masses of these black
holes appear to be correlated with the mass of their parent galaxy
(Wandel \& Mushotsky 1996), and the mass of the black hole is
typically $\sim 1$\% of the total bulge mass.

With the collision rate derived from the simulations and assuming a
mean mass increase of 10\,\msun\, per collision (which is consistent
with the $N$-body results) we can draw a relation between the mass of
the star cluster and that of the collision runaway.

\begin{figure}[ht]
\psfig{figure=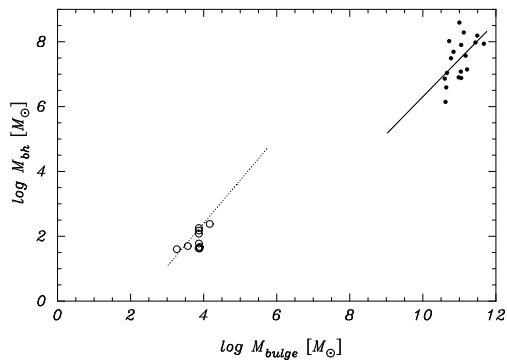,width=7.5cm,angle=-90}
\caption[]{The masses of the nuclear bulges and their measured black
holes (bullets, from Wandel 1999) and the results of a series of
$N$-body calculations (circles, PZHMM). The solid line gives the least
squares fit to the bullets, the dotted plots Eq.\,\ref{Eq:ncoll} from
the starting point given by the circles.
}
\label{fig:bhmass}
\end{figure}

This relation is presented as the dotted line in
Fig\,\ref{fig:bhmass}.  The solid line in Fig\,\ref{fig:bhmass} gives
a least squares fit to the measured black hole masses in galactic
bulges (Wandel 1999).  The horizontal offset may be explained by the
much smaller relaxation time of the simulated systems compared to that
of spheroidal bulges; galactic nuclei are much more massive than the
simulated clusters. Increasing the initial relaxation time in the
simulated clusters causes the dotted line in Fig.\,\ref{fig:bhmass} to
shift to the right.  The similarity in the slopes is quite striking
and may indicate that the black holes in galactic bulges are formed
via stellar collisions in the violent relaxation phase.

\section{Epilogue: Observing the models}

As the models become more elaborate and realistic, it becomes more and
more important to compare the results more directly with observations.
In a recent attempt to visualize the simulations, we started by
creating images of the models as if they had been observed through a
telescope.  The atmospheres' transparency as well as the response of
the CCD with realistic noise levels are modeled in great detail.

Comparisons between model data and the way an observer would study a
real cluster provides a critical check on the observer's data reduction
process.  In addition, there are many uncertainties as to how our
models would appear if they were observed as a real cluster on the
sky. Numerical problems, erroneous initial conditions, and other
incorrect assumptions may reveal themselves much more clearly when the
models are analyzed by observers rather than by theorists.  Since
theoretical models are at last producing sufficiently detailed data
that we can now talk meaningfully about performing comparisons between
theory and observations, it is critical that we identify robust ways
of projecting numerical data onto the observational plane. 

\begin{figure}[ht]
\hspace*{1.cm}
\psfig{figure=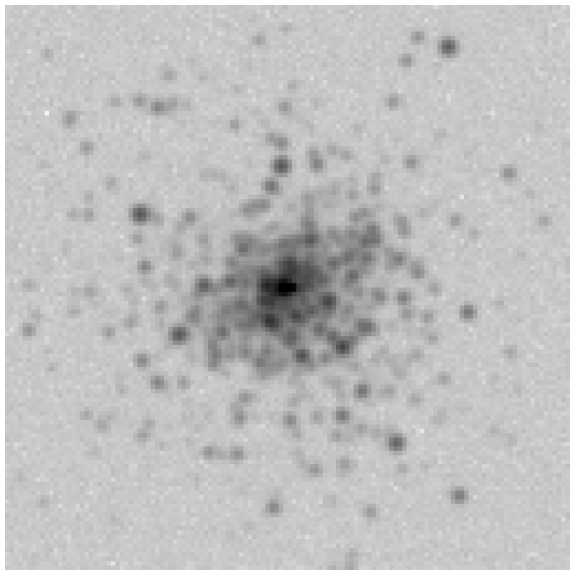,width=7.5cm,angle=-90}
\caption[]{Artificial CCD image of one of the simulated clusters from
a distance of 10 kpc and observed through HST in the equivalent $V$
band.  The simulated CCD images are available in FITS format in $B$ as
well as in $V$ via {\tt
http://www.sns.ias.edu/$\sim$starlab/research/project\_\,CCD}.  }
\label{fig:CCD}
\end{figure}

The main reason for doing this is to get an improved handle on how the
results of the numerical models would be viewed by an observer on or
in orbit around a distant planet.  An example of such an image is
presented in Fig.\,\ref{fig:CCD}.  More information about this project
is available at {\tt
http://www.sns.ias.edu/$\sim$starlab/research/project\_\,CCD}.

\acknowledgements I am grateful to Stephen McMillan, Jun Makino, Piet
Hut and Tereasa Brainerd for discussions.  Drexel University and the
University of Tokyo are deeply acknowledged for their hospitality and
the use of their GRAPE hardware.  This work was supported by NASA
through Hubble Fellowship grant HF-01112.01-98A awarded by
the Space Telescope Science Institute, which is operated by the
Association of Universities for Research in Astronomy, Inc., for NASA
under contract NAS\, 5-26555.


\section{Discussion}

{\bf Hans Zinnecker:}
The real 30\,Doradus cluster may even be more complicated
and interesting that you described, because initialy the proto cluster
also contains a lot of gas in which the stars have to move around
(see Bonnell, Bate \& Zinnecker, 1998, MNRAS 298, 93).

{\bf SPZ:} The presence of gas may indeed affect the dynamical
evolution of the system significantly, but it is not trivial to
include these effects self consistently in our calculations. At the
moment Piet Hut, Junichiro Makino, Stephen McMillan and I are working
on a series of similar $N$-body calculations in which we include a
large fraction of primordial binaries. These calculations include
stellar evolution but we neglect the primordial gas.

{\bf G\"oran \"Ostlin:}
Would a Wolf-Rayet star formed through a collision differ from an
ordinary single Wolf-Rayet star.

{\bf SPZ:} There are a number of ways in which a collision product
(also if it is a Wolf-Rayet star) can be distinguished.  A collision
producst is generally more massive and will appear younger than the
other stars in the cluster due to mixing during the merger event.
Such a star may manifest itself as a blue straggler.  Most collisions
are off center and the collision product is likely to be rapidly
rotating. This may show up in the spectra of the collision product.
Three stars in R\,136 (R136a3, Mk34 and R136-10) have unusual Hydrogen
abundances in their spectra (Massy \& Hunter, 1998) and it may worth
to study them in more detail.

{\bf Pavel Kroupa:} The central density decreases almost from the
start of your simulations. What is the cause of this.

{\bf SPZ:}
The cluster starts expanding after the first binaries have
formed, which is an extremly efficient process in the presence of a
mass function.  Note that the crossing time in these models is only
$\sim 80\,000$\,year, which is also the timescale on which the most
massive stars sink to the clusters center, whare they form hard
binaries. Binary formation and subsequent heating are the main source
for the expansion of the cluster.

\end{document}